\documentstyle[12pt]{article}

\begin{document}
\baselineskip = 20pt

\def\footnotefont{\tenpoint}
\def\figures{\centerline{{\bf Figure
Captions}}\medskip\parindent = 40pt%
\def\fig##1##2{\medskip\item{FIG.~##1.  }##2}}
\newwrite\ffile\global\newcount\figno \global\figno=1
\def\fig{fig.~\the\figno\nfig}
\def\nfig#1{\xdef#1{fig.~\the\figno}%
\writedef{#1\leftbracket fig.\noexpand~\the\figno}%
\ifnum\figno=1\immediate\openout\ffile=figs.tmp\fi\chardef\wfile=
\ffile%
\immediate\write\ffile{\noexpand\medskip\noexpand\item{Fig.\
\the\figno. }
\reflabeL{#1\hskip.55in}\pctsign}\global\advance\figno by1\findarg}

\parindent 25pt
\overfullrule=0pt
\tolerance=10000
\def\RR{${\rm R}\otimes{\rm R}~$}
\def\NSNS{${\rm NS}\otimes{\rm NS}~$}
\def\Re{\rm Re}
\def\Im{\rm Im}
\def\titlestyle#1{\par\begingroup \interlinepenalty=9999
     \fourteenpoint
   \noindent #1\par\endgroup }
\def\tr{{\rm tr}}
\def\Tr{{\rm Tr}}
\def\half{{\textstyle {1 \over 2}}}
\def\calt{{\cal T}}
\def\ie{{\it i.e.}}
\def\np{Nucl. Phys.}
\def\pl{Phys. Lett.}
\def\pr{Phys. Rev.}
\def\prl{Phys. Rev. Lett.}
\def\cmp{Comm. Math. Phys.}
\def\quart{{\textstyle {1 \over 4}}}
\baselineskip=14pt
\pagestyle{empty}
{\hfill DAMTP/96-??}
\vskip 0.4cm
\centerline{WORLD-VOLUMES AND STRING TARGET SPACES}
\vskip 1cm
 \centerline{ Michael B.  Green\footnote {M.B.Green@damtp.cam.ac.uk}
}
\vskip 0.3cm
\centerline{DAMTP, Silver Street,}
\centerline{Cambridge CB3 9EW, UK}
\vskip 1.4cm
\centerline{ABSTRACT}
\vskip 0.3cm
String duality suggests a fascinating juxtoposition of world-volume and
target-space dynamics.  This   is particularly apparent in the $D$-brane
description of stringy solitons that forms a major focus of this
article\footnote{This article is based on a  talk  presented at the CERN
Workshop on String Duality (December, 1995) and the published version of a
talk at the Buckow Symposium (September, 1995).}  (which  is {\it not}
intended to be a comprehensive review of this extensive
subject).  The article is divided into four sections:
\begin{itemize}
\item The oligarchy of string world-sheets
\item  $p$-branes and world-volumes
\item  World-sheets for world-volumes
\item  Boundary states,  $D$-branes and space-time supersymmetry
\end{itemize}

\vfill\eject
\pagestyle{plain}
\setcounter{page}{1}

\section{The oligarchy of string world-sheets}

According to current lore all  known superstring perturbation expansions may be
viewed as different approximations  to a single  underlying theory.  Viewed
from the standpoint of any given string theory the fundamental strings of other
theories are BPS solitons  that
are not apparent as particle states in string perturbation theory.  In addition
there are other solitonic solutions with $p$-dimensional spatial extension
which  play a vital r\^ole in the duality symmetries of the theory.  All of
these solitons have the property that they preserve some fraction of
space-time supersymmetry so that they are analogous to the BPS monopoles in the
Higgs--Yang--Mills system.  The  interrelations between different perturbation
expansions \cite{hulla,wittene} (and references therein)  appear to be profound
 and should indicate the path towards a more fundamental way of expressing the
theory in which no particular perturbative approximation has preferred status.

However, strings ($p=1$)  do  have an exalted position among this panoply of
solitons.   Only those theories in which the fundamental states are strings
possess  well-defined (albeit non-convergent) world-volume and target-space
perturbation expansions.   Thus, although first-quantized point-particle
($p=0$)  quantum mechanics is well-defined,  a theory including gravity cannot
be second-quantized in perturbation approximation.  Moreover, $p$-brane quantum
mechanics with $p>1$ is ill-defined --  the world-volume theories for such
$p$-branes are non-renormalizable quantum theories and can only be interpreted
as effective theories, suitable for describing long wavelength behaviour.

But,   following recent developments it has become plausible that the
$(p+1)$-dimensional world-volume field theories of $p$-branes should be
interpreted as effective low-energy approximations to an underlying string
theory.  This is implicit in the $D$-brane description of $p$-branes carrying
the charges of the Ramond--Ramond (\RR) sector  \cite{polchina,wittenb} (and a
multitude of other papers)  and plausibly (but more speculatively) of the
solitonic heterotic string and  fivebrane  \cite{kutasova} which carry
Neveu--Schwarz - Neveu--Schwarz (\NSNS) charges.

This represents the implementation of  the  idea   that the
string world-sheet of  one string theory (the \lq effective' theory) might be
interpreted as an  effective two-dimensional target-space of another string theory 
(the \lq underlying' theory) \cite{greenw}.  The  world-sheet coordinates of the  
effective theory are
world-sheet fields of the underlying theory while the world-sheet fields of the
effective  theory are
target-space fields of the underlying theory.  The string tension of the
effective
theory (which is the effective world-sheet loop counting parameter) is then
determined  in terms of the target-space closed-string coupling constant, $g$,
of the
underlying theory.  If the underlying string theory is a closed-string theory
the effective string tension behaves as $1/g^2$ while if it is an open-string
theory the effective string tension behaves as $1/g$.  In this way there is a
kind of  \lq world-sheet democracy' among string theories.

It is natural to extend this idea to the description of $p$-branes with the
($p+1$)-dimensional world-volume field  theory replaced by a corresponding
string theory.  Thus,  world-volume dynamics should be  well-defined in terms
of  the underlying string theory.  In this sense there is a \lq world-sheet
oligarchy' controlling the dynamics of $p$-branes.  Indeed,   in the $D$-brane
description of \RR $p$-branes  \cite{polchina,wittenb}  the underlying theory
is an open superstring theory and (more speculatively)  in the cases of the
heterotic  string and five-brane the underlying theory is type IIA closed
superstring theory   \cite{kutasova} (in which the world-volume fields are
twisted closed strings  pinned to an  orbifold point).

This leaves open the question of the relation of string theory to
eleven-dimensional supergravity -- or rather to $M$-theory \cite{horavaa},  a
Mythical eleven-dimensional quantum mechanical theory from which all known
string perturbation  theories can be obtained by compactification to lower
dimensions. $M$-theory should possess a  two-brane soliton as well as its
magnetic five-brane partner, which arise as classical solutions of
eleven-dimensional supergravity -- the \lq classical limit' of $M$-theory.
These solitons reduce to a variety of both   \NSNS
and \RR  $p$-branes in lower dimensions but it is currently a mystery as to how
the eleven-dimensional solitons may be expressed in terms of a  fundamental
quantum theory of strings.  The possibility that  a   twelve-dimensional
supersymmetric theory with $(10,2)$ signature -- \lq $F$-theory' -- should play
a r\^ole in these ideas has been illuminated in a  recent interesting paper
\cite{vafaa} that makes use of the properties of the  string theory with local
$N=2$ supersymmetry. It may be that $M$ \lq theory' and $F$ \lq theory'  can be
understood as some combination  of the ideas in \cite{greenw} and \cite{vafaa}.

Another issue that is related to $p$-brane dynamics   is the possible
existence of a new energy scale in string theory beyond the string scale
\cite{shenkerb}, which might be identified with the scale for exciting  higher
soliton modes.  This may also be connected to the presence of point-like
fixed-angle scattering at high energy in string perturbation theory in the
presence of Dirichlet boundaries \cite{greenold}.

The next  two sections  of this article  will  review  certain
features of $p$-branes, $D$-instantons \cite{polchinc,greenc} and $D$-branes in
order to  illustrate these points.   In  section 4  space-time
supersymmetry of $D$-branes and some of their interactions will be described
using methods that were originally applied to the case of  purely Dirichlet
boundary conditions
($p=-1$) \cite{greeng,greenc}.   Actually, the arguments apply to \lq
$(p+1)$-instantons' \cite{guta}.  These are related to $p$-branes by a double
Wick
rotation so that the world-volume is euclidean and  time is identified with
one of the directions transverse to the $(p+1)$-dimensional world-volume.

\section{ $p$-branes and world-volumes }

In the original  type IIA and IIB superstring  theories the fundamental
string states  carry  charges that couple to the  massless antisymmetric tensor
potential, $B^N_{\mu\nu}$,  of the  \NSNS sector.   The string-like soliton in
the  \NSNS sector of  either theory  is  a source for the \lq  electric' $B^N$
charge  while its \lq magnetic' partner  is a five-brane
solution carrying the magnetic $B^N$ charge.  [In general an \lq
electric' $p$-brane  charge is the integral of $*(F_{p+2})$ on the
$(8-p)$-sphere that bounds the space outside  the charged object in ten
dimensions while its
\lq magnetic' partner is carried by a $(6-p)$-brane which has a magnetic charge
given by the integral of $F_{p+2}$ on the $(p+2)$-sphere surrounding it.]

In addition to the two-form potential of the \NSNS\  sector  there are
several other massless ($p+1$)-form potentials  (with $(p+2)$-form field
strengths, $F_{p+2}$)  that arise as fundamental string states
of the \RR\ sector.  These comprise:  $a$ (which is the \RR\  pseudoscalar, or
zero-form),  $B^R_{\mu\nu}$ and  $A_{\mu\nu\rho\tau}$ in
the type IIB theory and  $A_\mu$ and
$A_{\mu\nu\rho}$ in the type IIA theory.  Thus, there
are  potentials of rank
$0,1,2,3,4$ in the \RR\ sectors of the type II  theories.
Although there are conserved charges associated with these potentials
there are no fundamental  string states that carry these charges.  This is
reflected  by
the fact that the potentials  do not couple minimally to the fundamental
string
world-sheet.  However,  $p$-brane solitons could  carry the
electric or magnetic \RR charges  with $p$
taking any of its possible values  from $p=-1$  to $p=9$.  Such
solutions of the type II  supergravity theories have indeed been found for all
values of $p$ in its range, $-1\le p \le9$.
The zero-brane (black hole), two-brane, four-brane, six-brane and
eight-brane solutions of the type IIA
theory and  the one-brane
(string), self-dual three-brane, five-brane and seven-brane solitons of
type IIB string are standard solitons (many of these are reviewed in
\cite{duffa}).

There is  actually an infinite number of  \lq dyonic' \RR one-branes in the
type IIB theory carrying (coprime) integer values of the charges that couple to
 the antisymmetric tensors in the \NSNS and \RR sectors \cite{schwarzx}.
Likewise, there is an infinite set of five-branes carrying the magnetic
charges.  The case $p=9$ is  special since the accompanying field strength
vanishes identically, and the $10$-form potential itself  is connected with
the presence of chiral anomalies in
type I theories with any gauge group other than $SO(32)$. The $p=8$ soliton
couples to a cosmological constant in the type IIA
theory \cite{polchina,polchinb,bergshoeffa} and is a solution of \lq
massive' type IIA  supergravity \cite{romansa}.

The world-volume field theory that defines  the classical dynamics of each of
these $p$-branes is a $(p+1)$-dimensional  field theory with extended
supersymmetry.  For $p>2$ this necessarily involves world-sheet fields with
spins greater than $1/2$ -- for example,  a  brane  of the \RR sector is
described by a  world-volume theory that is a supersymmetric abelian gauge
theory \cite{calstroma,dufflu} in which one of the fields is the spin-1 photon.
 This world-volume  gauge potential emerges from zero modes of the
ten-dimensional theory fields in the background of a brane,  whereas in the
$D$-brane picture it is identified with a  ground state of the open
superstring.

The case of the $-1$-brane deserves special attention  since it is an
instanton rather than a solitonic particle.  Instead of carrying a
charge, such a
solution describes a transition which leads to the non-conservation of the
charge to which it couples.  This is the Noether charge  associated
with the  translation symmetry of the \RR\ scalar potential, $a$, in
the classical theory.  Unlike
other \RR\ charges this is a global charge and its violation is to be
expected in string theory.  In fact, the instanton solution \cite{gibbonsa}
has
the string-frame  interpretation of a space-time Einstein--Rosen wormhole
\cite{einstein}
connecting two asymptotically flat regions.   Particles carrying the charge
can fall through the neck connecting one region to the other which leads to its
non-conservation.

The fact that these solutions  preserve some fraction of the maximal
original supersymmetries means that they  have
much more tractable quantum corrections than generic,
non-supersymmetric, solutions and it follows that  certain statements about the
strong coupling limit are reliable.  The prototype for this  is the
Montonen--Olive  conjecture \cite{montonena,olivea,osborna} relating weak and
strong coupling
limits of
$N=4$ supersymmetric Yang--Mills theory generalized to a conjectured
$SL(2,Z)$ symmetry of the theory \cite{sena}.    The field theories to which
this conjecture  and its generalizations  apply may be viewed as low
energy limits of superstring theories and this  suggests a generalization of
the conjecture to the string setting.   There is by now much evidence
for these stringy non-perturbative dualities.

\section{World-sheets for world-volumes}

\noindent{{\it $D$-branes and $D$-instantons}\hfill\break
The $p$-brane solitons carrying charges in the \RR sector of the low energy
supersymmetric field theory have an intrinsically stringy
manifestation.   These stringy versions of branes, known collectively as
$D$-branes \cite{daia,leigha,polchina} (or $D$-instantons when  $p=-1$
\cite{polchinc}),    are associated with {\it open} superstring configurations.

The essential point is
that a  solitonic solution of perturbative field theory has
a mass (more properly, a mass per unit volume)  that becomes infinite in the
limit of zero coupling, $g\to 0$  --
typically the mass behaves as $1/g^2$ or (in the case of the solitons of
the \RR\ sector) $1/g$.   Perturbation theory in the presence of a soliton is
not manifestly  invariant under  ten-dimensional translation invariance.   So
if the brane
world-volume
is taken to be a flat lorentzian  sheet in $p+1$ dimensions translation
invariance is spontaneously broken  in the
$(9-p)$ transverse dimensions and is only restored after integration over
translational zero modes.   There are
interesting open superstring configurations which  describe fluctuations around
just
this kind of background. These are open superstrings with their
end-points fixed in  the world-volume of the brane -- in other words in the
directions labelled by  $i=  p+1, \dots, 9$ the coordinates satisfy
$\partial_t X^i =0$ (where $t$ denotes the derivative tangential to
the boundary).  In the  other directions, $X^{\alpha}$ with
$\alpha =
0, \dots, p$,   the end-points satisfy the usual open-string Neumann
boundary conditions, $\partial_n X^\alpha=0$ (where $n$ denotes the normal
derivative at a
world-sheet  boundary).   These configurations  preserve  half of the
space-time supersymmetries -- they are stringy analogues of BPS
states.
So we see that whereas the standard perturbative states of type II superstring
theories are closed-string configurations,  in the presence of $p$-brane
solitons there are additional open-string states in the theory with end-points
moving in the brane.  These open-string configurations  owe their existence to
the presence of the brane to which they are tethered.

The coupling of these  open  superstrings to the brane give rise to
fluctuations
that define the brane dynamics.  Thus,  a pair of string end-points may be
created   in the world-volume of the brane at some time and  annihilate at a
subsequent time.  In between these times the end-points trace out  a closed
curve in the world-volume.  The intermediate  open-string histories define a
world-sheet that is a disk bounded by this curve.  This means that the boundary
conditions on the disk are  Neumann in $p+1$ dimensions and
 Dirichlet in $9-p$ dimensions.   The functional integral over the disk gives
the leading contribution to the ground-state
energy of the $D$-brane  that is of order
$c /g \sim c e^{-<\phi>}$ where $c$ is a known constant  \cite{polchinc}.
The factor of $1/g$, which here arises from the fact that the Euler character
of a disk is -1,  is  characteristic of the energy of  solitons in the  \RR
sector.

Independent fluctuations of the brane can arise in disconnected
regions of the world-volume, described by independent disk world-sheets.  The
case of the $D$-instanton is special since the world-sheet
boundaries are fixed at a position $y^\mu$ ($\mu =0,\dots,9$) \cite{polchinc}
which is
a constant.   Since  the  boundaries of the disks are all at the same
space-time
point  momentum is conserved  after integration over the
space-time position of the instanton.  With $p>-1$ momentum is not conserved
locally  unless account is taken of the excitation of long
wavelength  modes of the brane.  These modes  carry the momentum
between the disconnected  regions spanned by the separate  disks.

The full string partition function involves a sum over all possible numbers of
such
fluctuations and this gives rise to an exponential of the individual
disk diagram  in the functional integral.
In addition to these iterations of disconnected disk diagrams there
are iterations of  diagrams
with more complicated topology that also contribute to the dynamics of
a single $D$-brane.   The simplest of these is the  diagram in which
the boundaries of
two disks form the ends of a cylindrical world-sheet.  This describes a
higher-order vacuum  effect in which a closed string is exchanged between
the two boundaries.  Multiple  iterations of this diagram  must be
included in the functional integral  and
these also exponentiate.  The general diagram contributing to a single brane
consists of the exponential of connected world-sheets with arbitrary numbers of
boundaries
and handles.

For example, in the background  of a single $D$-instanton  the sum over all
diagrams contributing to the  vacuum functional  is given by
\begin{equation}\label{singinst}
Z =\int d^{10} ye^{S^{(1)}}.
\end{equation}
In this expression the single $D$-instanton action,  $S^{(1)}$ , is given by
\begin{equation}\label{oneinstact}
S^{(1)} = \sum_{r=0}^\infty g^{r-2} f_r(y),
\end{equation}
where $f_r$ is the functional
integral over {\it connected} world-sheets with $r$ boundaries (and
there is also a sum over handles that has been suppressed) \cite{greenc}.
The background
fields are required to satisfy the
conditions that ensure conformal invariance and the fluctuations of these
fields determine the perturbative vertex operators in the usual
manner.   In the absence of fundamental external closed-string states the
$r=0$ term (the spherical world-sheet) vanishes by the usual scaling argument
and the terms with $r>1$ vanish by supersymmetry, leaving simply the disk
diagram of order $1/g$  in the exponent of (\ref{singinst}).     The instanton
 therefore generates non-perturbative contributions to the scattering
amplitudes of the form  $e^{c(ia_0 -1/g)}$ (where $a_0$ is the constant value
of the \RR scalar field that is a parameter of the instanton solution).   The
occurrence of such effects in string
theory  was anticipated by Shenker \cite{shenkera} by considering  the
rate of divergence of string perturbation theory and is strikingly
different from soliton and instanton effects in standard field theories
such as QCD which are typically of order $e^{-c/g^2}$.

  The generalization of (\ref{singinst})   to the situation in which there are
arbitrarily many  BPS (and no anti-BPS) $D$-instantons is straightforward
\cite{greenc}.
\vskip 0.2 cm

\noindent{\it Scattering in the presence of  $D$-instantons and
$D$-branes}\hfill\break
Fundamental closed-string states scatter  from  a $D$-brane by interacting with
the  open-string fluctuations.  The amplitude is given in leading approximation
by attaching
closed-string vertex operators for the fundamental string states  to the disk
world-sheet.  If  the brane is fixed in space at a
constant transverse position $y^i$ ($i=p+1,\dots, 9$)  momentum is not
conserved in the process but once the
zero modes of the $D$-brane are taken into account (which include the
overall translation mode) momentum conservation will be restored.
In general, there are bosonic zero modes associated with local
transverse fluctuations of the brane so that the position of the brane
is the field, $Y^i(X^\alpha)$.   According to \cite{polchina} and
\cite{wittenb} infinitesimal fluctuations are induced by coupling the
open-string massless vector potential to the boundary of the disk.
This vertex operator has the form $\oint d\sigma A_i(X^\alpha)
\partial_n X^i$, where $A_i(X^\alpha)= Y^i(X^\alpha)  -y^i$ are the
components of the vector potential transverse to the brane at the
point $X^\alpha$.  The brane can be given a uniform velocity  by  choosing
$A^i$ to be linear in $X^0$ \cite{bachasa,klebana}. The other components of the
vector potential, $A_\alpha(X^\beta)$,
comprise a world-volume vector  field.   In addition, the fermionic zero modes
associated with the broken supersymmetries are excited by coupling to the
massless open-string fermion vertex.

Scattering amplitudes of fundamental closed-string states obtained from
(\ref{singinst}) by considering small fluctuations of the bulk background
fields consist of sums of functional integrals over both connected and
disconnected world-sheets,  rather than simply  the connected world-sheets as
in conventional (Neumann)  open-string
theory.   The exponentiation of  world-sheets plays an
essential r\^ole in ensuring the cancellation of  the novel Dirichlet
divergences that
arise when considering individual connected world-sheets
\cite{polchinc,greenc}.

Scattering amplitudes in the background of a $D$-instanton exhibit point-like
fixed angle behaviour order by order in perturbation theory around the
instanton \cite{greenc}.  In fact this striking feature was one of the main
reasons for the original interest in string theory with fully Dirichlet
boundary conditions  \cite{greenold} (and contrasts with the exponential
decrease of fixed-angle scattering cross sections in closed-string theory
\cite{amatia,mendea}).  Although originally studied in the bosonic theory this
point-like behaviour is also a feature of  fixed-angle superstring amplitudes
in  perturbation theory in the  background of a $D$-instanton \cite{gutb}.  A
$D$-brane is not point-like -- this is a simple consequence of the Neumann
boundary condition in the time-like direction  \cite{gutb,kleba,klebb}.
However, the length scale that characterizes its interactions order by order in
perturbation theory, $\sqrt \alpha'$,  does not take into account the excited
modes of the brane.  Since the effective tension of a \RR brane is of order
$1/g \alpha'$ a  new  momentum scale should arise  beyond which high frequency
modes on the brane can be excited.  The possible existence of such an
intrinsically non-perturbative  scale seems related to observations in
\cite{shenkerb}.
\vskip 0.2cm

\noindent{\it Effective world-volume theories}\hfill\break
The action of a  $D$-instanton is particularly simple because all boundaries
are mapped to the same point in the target space-time and (\ref{singinst}) is
simply an integral over the position of that point.
With a single $D$-brane of $p \ge 0$ spatial dimensions the integration over
$y^i$
in (\ref{singinst})  is (at first sight) replaced by a functional
integration over the values of the world-volume fields $Y^i(x^\alpha),
A_\alpha (x^\alpha)$  (and their fermionic partners) on the world-sheet
boundaries.  The resultant
vacuum functional can be argued to  have the form
\begin{equation}\label{singpbrane}
Z =\int D^{9-p} Y^i(x^\alpha) D^{p+1} A_\beta(x^\alpha) e^{S^{(1)}[Y,A,
\cdots]},
\end{equation}
where $S^{(1)}[Y, A, \cdots]$ is the ($p+1$)-dimensional world-volume action
functional and
the functional integral is over the fields living in this world-volume (the
dots indicate the fermionic fields  and dependence on the background fields is
implicit).  This action is itself  the low-energy limit of  a  sum of
conventional string functional integrals (over $X^\mu(z)$) on connected
world-sheets with boundaries coupling to the massless open-string fields -- the
 abelian Yang--Mills  potential $A_\mu = (Y^i,A_\alpha)$  and its massless
open-string fermionic partners --  and on which the boundary conditions are
Dirichlet in the directions labelled, $i$.      Since the effective action
originates from the ten-dimensional open string theory it possesses half the
number of world-volume supersymmetries that the  original type II theory has.
This naturally agrees with the fact that  a $p$-brane is a BPS soliton in the
presence of which half the  supersymmetry is spontaneously broken.

For example, in the case of a single $D$-string ($p=1$)  the effective action
can be argued to
have  the Nambu--Goto/Born--Infeld form \cite{leigha,schmidhubera,townsenda},
\begin{equation}\label{effact}
 S^{(1)}[Y, A, \cdots]  =  {-1\over 2 } \int d^2 x  \left(e^{-\phi} \sqrt{-
\det (G + {\cal F})} \right.   \left. +\half \epsilon^{\alpha\beta}
B^R_{\alpha\beta} - \half  a
\epsilon^{\alpha\beta} {\cal F}_{\alpha\beta}    + \dots + O(g^0)\right),
\end{equation}
where the effective world-volume coordinates, $x^\alpha$, are the zero modes of
the original world-sheet fields and  where  ${\cal F}_{\alpha\beta} =
F_{\alpha\beta} -B^{N}_{\alpha\beta}$ and $G_{\alpha\beta}$ is the induced
metric on the effective  world-sheet.   $F_{\alpha\beta} = \partial_{[\alpha }
A_{\beta]}$ is the Maxwell field strength in the world volume and  the dots
indicate   terms involving
fermionic world-volume fields that  complete the $N=8$ supersymmetry of the
world-sheet action (these fermions on the effective world-sheet are components
of the dimensionally reduced fermion of ten-dimensional supersymmetric Maxwell
theory).

This effective Born--Infeld action may be motivated as in \cite{schmidhubera}
by considering the modification of the string tree-level $\beta$ function
conditions  in the presence of a constant open-string field strength --
generalizing the discussion of the purely Neumann theory
in\cite{fischlera,callana,polchinq}.  In fact, the Born--Infeld action has long
been known to arise from the open-string theory \cite{neveua} and
\cite{fradkina}.
 For fluctuations of the string around its straight configuration the induced
metric  is given  by $G_{\alpha\beta} = \eta_{\alpha\beta} + \partial_\alpha
Y^i \partial_\beta Y^i$ and the bosonic part of the Wess-Zumino term becomes
$B^R_{\alpha\beta} + B^R_{ij}\partial_\alpha Y^i \partial_\beta Y^j$.
  The leading term in the exponent in (\ref{singpbrane}) is again
determined by the disk diagram while  higher-order terms arise  from connected
world-sheets of higher genus.       In  \cite{schmidhubera} it was shown
that the full dyonic spectrum of type IIB string solitons with tensions
$T_{m,n} = \sqrt{(na-m)^2 + n^2/g^2}$ (where $m$, $n$ label the \NSNS and \RR
antisymmetric tensor charges, respectively)  \cite{schwarzx} can be summarized
in the form  (\ref{effact})  by considering $n$
coincident $D$-strings with background ${\cal F}$.  This is the spectrum
expected from the arguments in \cite{wittenb}, although the delicate issue of
why the only  stable states  are those with $m$ and $n$ coprime requires a more
subtle argument.

 With  $n$  parallel $D$-branes  there  are $n$ independent $U(1)$ gauge
potentials associated with the ground states of open strings with both
end-points on the same brane.  These potentials may be denoted $A_\mu^{rr} =
(A_\alpha^{rr}, A_i^{rr})$ where $r=1,\cdots,n$ labels the brane (and $A_i^{rr}
= Y_r^i - y_i^r$ defines its transverse position).  There are also  $n^2 -n$
 gauge potentials associated with ground states of open strings with end-points
on different branes, $A_\mu^{rs} = (A_\alpha)^{rs}, A_i^{rs})$ ($r\ne s$).
When the branes  are separated   the extra potentials are massive but they
become massless in the limit of coincident branes --  when all the branes are
coincident there is a nonabelian  $U(n)$ gauge symmetry
\cite{polchina,wittenb}.    The dynamics of the system of parallel branes is
thus related to  spontaneously broken supersymmetric $U(n)$ Yang--Mills theory
(with Yang--Mills potentials, $A_\alpha^{rs}$) compactified from 10 dimensions
to $p+1$ dimensions where the scale of the breaking is determined by the
separation of the branes.   Very interestingly,  since the spatial  coordinates
$Y_i^r$ are also components of the ten-dimensional gauge potentials they also
form part of a larger non-commutative algebra when the branes are closely
spaced, as  was emphasized in \cite{wittenb}.  This  is reminiscent of ideas in
\cite{connesa}.

At second sight, however, (\ref{singpbrane}) does not lead to a  well-defined
world-volume quantum theory since the theory with action $S^{(1)}$ is
non-renormalizable.  The world-volume theory should really be defined by the
complete  open superstring theory compactified to $p+1$
dimensions, rather than by its low-energy approximation.  In that case
(\ref{singpbrane}) would be an integral over open-string fields.  This bears
some resemblence to the suggestion, made for the case  $p=1$ in \cite{greenw},
that the world-volume  manifold might be considered to be an approximation to
the ($p+1$)-dimensional target space of a string theory.  The case $p=1$ is an
example of the idea that a string world-sheet might be described by the
target space of another string theory.

To summarize:-
\begin{itemize}
\item   The classical  world-volume theory of a  \RR $p$-brane is
obtained in the $D$-brane description as a low-energy limit of  ten-dimensional
open-string theory with  Neumann boundary conditions   in $p+1$ directions and
Dirichlet conditions in the remaining $9-p$ directions.  The ten-dimensional
massless open-string ground state  fields are interpreted as world-volume
fields of the brane.  Although
the world-volume action may  not be a well-defined $p$-brane  quantum theory
the full open-string theory is.   The fact that the open-string loop expansion
is a power series in $g$ implies that the $p$-brane tension behaves like $1/g$
for small $g$.
\end{itemize}
The idea that $p$-branes are defined by an underlying string theory plausibly
also applies  to the $p$-branes of the \NSNS sector, such as the heterotic
one-brane and five-brane:-
\begin{itemize}
\item   The world-sheet action of the heterotic string (or $1$-brane) may be
obtained
as the low-energy limit of the  compactification of ten-dimensional type IIA
theory on an
orbifold, $K_3 \times T^4/Z_2 \times R^2$, where $Z_2$ acts on
non-compact coordinates \cite{kutasova}.   Furthermore, the \NSNS fivebrane
soliton of the type IIA theory  has a world-volume action that is the
low-energy limit of the compactification of the type IIA theory on an orbifold,
  $T^4/Z_2
\times R^6$.  Again, the underlying string theory gives a well-defined quantum
theory even if the field theory is not.    In these cases the target-space loop
expansion is a power series in $g^2$ since the underlying theory is a
closed-string theory so that the \NSNS $p$-brane tension behaves as $1/g^2$.
\end{itemize}
In both  cases the classical  world-volume fields are those that come from
strings that are tied to the brane.  In the case of  $D$-branes these are the
open-string fields while in the heterotic examples they are in the twisted
closed-string sector.   These are non-gravitational fields   so the classical
effective world-sheet theories are non-gravitational.  However, in either case
quantum loop corrections of the underlying world-volume string theory
necessarily induce gravitons which move in the ten-dimensional target space so
that the distinction between the world-volume and the target space will
disappear in the quantum theory.  This relationship between the world-volume
theory and the underlying string theory can be summarized by the following
table,
\vskip 1.2cm
\noindent
\begin{tabular}{|c|c|c|}
EFFECTIVE WORLD-VOLUME  & & UNDERLYING WORLD-SHEET \\
\hline && \\
 World-volume coordinates & $x^\alpha$ & World-sheet fields  \\
  && \\
 World-volume fields  &  $A^i \equiv Y^i$, $A^\alpha$ & Target-space fields  \\
&& \\
 World-volume loop-counting parameter & $e^{\langle \phi\rangle}$ &
Target-space loop-counting parameter  \\
\RR sector (tension)$^{-1}$  & $g$ & open string loop\\
\NSNS sector (tension)$^{-1}$ & $g^2$ & closed string loop\\
\hline
\end{tabular}
\vskip 0.5cm

More generally,
this  \lq confusion' between target space and world-volume is also illustrated
by the profusion of r\^oles for the same effective action -- it may describe a
world-volume from one point of view or a target space from another.  For
example:-
\begin{itemize}
\item   A two-brane soliton of  eleven-dimensional $M$-theory can have
a (string-like)  boundary that lives in a five-brane soliton
\cite{townsenda,stromingerb,douglasa}.  This self-dual string is described by a
world-sheet embedded in the six-dimensional world-volume of the
five-brane which itself is embedded in eleven space-time  dimensions.  The
five-brane solution breaks half the eleven-dimensional supersymmetry and the
embedded string soliton breaks half of the remaining supersymmetry.
On the other hand the same self-dual string theory arises directly in
six-dimensional space-time by the compactification of the type IIA
theory from ten dimensions on $K_3$ \cite{dufflub,wittenf}.
\end{itemize}

\section{Boundary states,  $D$-branes and space-time supersymmetry}

The energy between two parallel $D$-branes is determined to lowest
order  by the cylinder diagram
in which one boundary is fixed in the world-volume of one of the branes (at
$y_1^i$, say) while the other is fixed in the other brane (at $y_2^i$).
The expression for this diagram is given by
\begin{equation}\label{cyldia}
\sum_S  \int d\tau   \langle B^{(S)}, \eta_1,y_1|
e^{-H_S^{cl} \tau }P_{GSO} |B^{(S)}, \eta_2,
y_2\rangle,
\end{equation}
where the sum is over the \NSNS and \RR closed-string sectors
(labelled by $S$) and $H_S^{cl}$ is the usual closed-string hamiltonian
in the $S$ sector.  A sum over spin
structures is implied by the GSO projection in the cylinder channel
$P_{GSO}$.  The end-states satisfy the world-sheet supersymmetry
conditions \cite{greeng,polchina},
\begin{equation}\label{bpsstate}
(F^{(S)} + i \eta \tilde F^{(S)})|B^{(S)}, \eta, y\rangle =0,
\end{equation}
where $F^{(S)}(\sigma)$, $\tilde F^{(S)}(\sigma)$ are the left-moving
and right-moving
world-sheet supercharges.  The sign  $\eta =\pm 1$  determines whether
the state is a BPS state or an
anti-BPS state.  In the case that both end-states are of the same kind the
sum over spin structures vanishes by Jacobi's abstruse identity which
indicates that the energy  vanishes due to a  cancellation of the exchange of
states in the \RR and the \NSNS
sectors \cite{greeng,polchina}.

More generally, the interaction of two BPS $D$-branes with $M$  fundamental
string states is determined by the cylinder diagram with vertex operators
attached and is non-zero when $M\ge 2$.  The arguments of \cite{greene,greeng}
(and
references therein) show that in the case of $D$-instantons  this process
can be expressed in terms of position-space singularities on and inside the
light-cone.

 However, the correlation function of two boundary states of the opposite
type --  one BPS ($\eta_1=1$) and one anti-BPS ($\eta_2=-1$) --  does
not vanish.   This is due to a change in sign of one of the spin structures
in the \RR sector (the one that is  anti-symmetric along the axis of the
cylinder),
which breaks the supersymmetry completely.  This is analogous to the fact
that the BPS  monopole - anti-monopole system is not a BPS state and not an
exact solution to the equations of motion.  The absence of supersymmetry in
the correlation function of two boundary states of the opposite type is
reflected in the $D$-instanton case by the  presence of a singularity {\it
outside} the light cone --  at $(y_2-y_1)^2 = \pi \alpha'/2$ --  in
this process, which makes the correlation function ill-defined \cite{greeng}.
 Correspondingly, the correlation between two $D$-brane boundary states of
opposite type is badly behaved at separations smaller than the string scale
\cite{greenwil,susskinda}, at $({\bf y}_2 - {\bf y}_1)^2 = \pi \alpha'/2$
(where ${\bf y} = y^i$).

The fact that  $D$-branes preserve half the space-time
supersymmetry can be  seen from a  light-cone gauge  argument that was
originally applied to the case of  $D$-instantons
in \cite{greeng} but  which generalizes to the case of $D$-branes with $p\le 7$
 as in
\cite{guta}.  In the
light-cone gauge the directions in the  string  world-sheet are  identified
with two of the target-space directions, $X^\pm= X^0\pm X^9$. The time-like
world-sheet
coordinate is identified with the light-cone time, $X^+ = p^+ \tau$,  while
the the other world-sheet coordinate, $\sigma$, is chosen so that
$P^+(\sigma) =  p^+/2\pi$. Now consider a process described by a
world-sheet that is a semi-infinite cylinder spanned by an incoming
physical closed-string state (at $X^+ = p^+ \tau = -\infty$) evolving to
a boundary  end-state, $|B,y\rangle$ (the
parameter $\eta$ will be suppressed for now) at  $ X^+ = p^+ \tau = y^+$.
Since the boundary is at a fixed value of $X^+$ there is  at least one
direction in which the boundary condition is Dirichlet.  Furthermore, in
the light-cone gauge the  coordinate $X^-(\sigma)$ is determined by
$p^+\partial_\sigma X^- = \int_0^\sigma d\sigma' \partial_\tau
X^I\partial_{\sigma'} X^I + $ fermion terms  (where $I=1, \dots, 8$).   It
follows that  $\partial_\sigma X^- |B\rangle =0$  whether the
transverse coordinates satisfy Neumann or Dirichlet boundary conditions so that
 $X^-$ also
satisfies a Dirichlet condition.  This Dirichlet condition on $X^-$  is
consistent with the  non-conservation  of the conjugate momentum, $p^+$,  at
the boundary in the
process under consideration.  Thus,
both $X^+$  and $X^-$  satisfy
Dirichlet conditions while the transverse coordinates $X^I$ may satisfy
either  Neumann conditions, $\partial
X^\alpha/\partial\tau |B,y\rangle =0$ ($\alpha = 1, \dots, p+1$), or Dirichlet
conditions, $\partial X^i /\partial \sigma |B,y\rangle = 0$ ($i = p+1,
\dots, 8$).  Therefore, the following  argument will only apply to the case
in which $p\le 7$.  Furthermore, in this  situation   the $p+1$ Neumann
directions are euclidean directions transverse to the
world-sheet and the time direction is one of the $9-p$ Dirichlet directions.
This choice of  coordinates  does not describe a $D$-brane soliton but
rather an analytic continuation in which time is one of the coordinates
transverse to the ($p+1$)-dimensional euclidean world-volume.  Here, the
$p$-brane is viewed as an  instanton with action density localized  in a region
with
$(p+1)$-dimensional extension.  For example,  in the case $p=5$ the euclidean
theory in the space transverse to the five-brane can be interpreted as the
euclidean continuation of the four-dimensional \lq axionic' instanton,
which has extension in the six dimensions (now taken to be
euclidean) of the string world-volume (this interpretation has been used in
the field theoretic context in  \cite{giddingsa}).   The total action will only
be finite if the $p+1$  Neumann dimensions are compactified on a torus.

A  sixteen-component  ten-dimensional  supercharge  decompose in the light-cone
gauge
into
two inequivalent $SO(8)$ spinors, $Q^a$ and $Q^{\dot a}$  ($a =1,\cdots,8$).
 In the type II theories there are two such supercharges.  The
left-moving supercharges  satisfy the superalgebra
\begin{equation}\label{susyalg}
\{Q^a,Q^b\} =  \delta^{ab}p^+, \quad
\{Q^{\dot{a}},Q^{\dot{b}}\} = \delta^{\dot{a}\dot{b}}H, \quad
\{Q^a,Q^{\dot{a}}\}  = \gamma^I_{a\dot{a}}p^I.
\end{equation}
and a similar set of relations applies to  the right-moving supercharges.   In
this expression $p^I$ is the
zero mode of the transverse momentum, $P^I$,  and   $H = P^-$ is the light-cone
hamiltonian.
The type IIA theory has left-moving and right-moving 16-component
supercharges of the opposite chirality.  In that case  the role of the
dotted and undotted $SO(8)$ spinors
is interchanged between the left-moving and right-moving light-cone
supercoordinates.

A  Dirichlet boundary state that  preserves half the space-time
supersymmetry was constructed  for the case of the purely Dirichlet theory in
\cite{greeng} and for general $p$ in \cite{guta}  -- this should be viewed as a
manifestly space-time supersymmetric version of the state in \cite{polchina}
that preserves half the world-sheet supersymmetry.  The state is defined to
satisfy the conditions,
\begin{eqnarray}\label{dircos}
(\partial+\tilde{\partial})X^i| B, \eta\rangle & =& 0,  \qquad
i=1,\cdots,p+1 \\
(\partial-\tilde{\partial})X^i | B, \eta\rangle & =& 0, \qquad
i=p+2,\cdots,8,
\end{eqnarray}
on the bosonic coordinates and (in the case of the type IIB theory),
\begin{equation}\label{fermdir}
(S_n^a+ i \eta  M_{ab}\tilde{S}^b_{-n})\mid
B, \eta \rangle=0,
\end{equation}
on the fermionic coordinates (where $S^a$ and $\tilde S^a$ are left-moving and
right-moving light-cone gauge
spin fields that are the fermionic target-space spinors of light-cone
superstring
theory).
The matrix $M_{ab}$ is given by
\begin{equation}\label{mdef}
M_{ab} =  (\gamma^1 \cdots \gamma^{p+1})_{ab},
\end{equation}
which reduces to  $M_{ab}= \delta_{ab}$ for the case of the $D$-instanton and
guarantees
the  linear relations between the supercharges,
\begin{equation}\label{qcons}
 (Q^a  +i\eta
M_{ab}\tilde{Q}^b)\mid B,
\eta\rangle  = 0 , \qquad
(Q^{\dot{a}} + i\eta  M_{\dot{a}\dot{b}}\tilde{Q}^{\dot{b}}_{-n})|B, \eta
\rangle = 0.
\end{equation}
 The type IIA case has a similar expression (in which the matrix $M$ is the
product of  an odd number of $\gamma$ matrices).
These linear relations between the left-moving and right-moving space-time
supercharges are consistent with the superalgebra and thus, one half of the
space-time supersymmetry is preserved for any value
of $p$,  the case $p=-1$ (the $D$-instanton) coinciding with \cite{greeng}.  As
expected, the conditions for  $p>-1$ \cite{lia,guta} are not
Lorentz invariant
since the boundary state is associated with the presence of a soliton
background.

The supersymmetry conditions (\ref{qcons}) can be expressed in light-cone
superspace by introducing the $SO(8)$ spinor Grassmann coordinates,
$\theta^a = Q^a -  i M_{ab} \tilde Q^b$.  The general  boundary state is then a
superfield that has a component expansion in powers of $\theta^a$.
Imposing the conditions  (\ref{qcons})  makes all components  vanish  apart
from the bottom (for $\eta =1$) or top
 (for $\eta = -1$) one.  For example, for $p=-1$ the bottom
component  of the superfield is a  complex  combination  of the two
scalar
fields, $\phi$ and $a$ (the dilaton and the \RR\ scalar) and the top
component is the complex conjugate.  With  $\eta=1$
all components vanish  apart from the bottom one and the boundary
state  can be written in the form,
\begin{equation}\label{nonzero}
 | B\rangle = \exp\sum_{n=1}^\infty \left({-1\over n} \alpha_n^{i\dagger }
\tilde\alpha_n^{i\dagger } + i S_n^{a\dagger} \tilde S_n^{a\dagger}\right) |
B_0
\rangle,
\end{equation}
where the  zero-mode piece is given by
\begin{equation}\label{minone}
|B_0\rangle =   |i\rangle |\tilde i\rangle + i |\dot
a\rangle |\tilde{\dot a}\rangle .
\end{equation}
The states $|i\rangle$,  $|\dot a\rangle$  are $SO(8)$ vector and spinor
ground states in the left-moving sector and  $\alpha^{i\dagger}_n$,
$S^{a\dagger}_n$ are the creation modes of the bosonic and fermionic
coordinates  (and $\tilde{} $ denotes the right-moving sector).  This boundary
state  couples equally to
$\phi$ and $ia$ and is therefore an equal
source for the two scalar fields, which agrees with the field-theoretic
treatment of the $D$-instanton in \cite{gibbonsa} (where $\alpha =ia$ is the
\RR scalar in euclidean space).  These statements generalize to
arbitrary $p$, thereby reproducing (in linearized approximation) the
relationships that give rise to the BPS states of the low energy field
theory.

The basic $p$-branes are sources of  \RR\  $(p+1)$-form charges.    More
generally they are also sources of the \NSNS charges associated with $B^N$ when
there is a  non-trivial boundary condensate   of the open-string Maxwell field,
$A_\alpha$.  Such branes are also sources of  $r$-form \RR charges with $r =
p-1, p-3, \cdots$ \cite{wittenb}.  This is again a manifest consequence of  the
supersymmetric  description of   boundary states in the presence of
open-string  condensates \cite{guta} which resemble the convariant boundary
states \cite{lia}.
\vskip 0.2cm

\noindent{\it Acknowledgment}\hfill\break
I am very grateful to Michael Gutperle for many insights  and to David Kutasov
and Adam Schwimmer for a discussion related to  parts of section 3.

\end{document}